\documentclass[11pt]{IEEEtran2e} %
\pdfoutput=1
\usepackage{amsfonts}
\usepackage{amssymb}
\usepackage{url}
\usepackage{amsmath}
\usepackage{colordvi}
\usepackage{color}
\usepackage{balance}
\usepackage{graphicx}
\usepackage{latexsym}
\usepackage{balance}
\usepackage{cite}

\usepackage{times}

\newcommand{\urlBiBTeX}[1]{\url{#1}}
\newcommand{\urlbibteX}[1]{\url{#1}}
\def\BibTeX{{\rm B\kern-.05em{\sc i\kern-.025em b}\kern-.08em
    T\kern-.1667em\lower.7ex\hbox{E}\kern-.125emX}}

\def\conv{*}
\def\deconv{\oslash}
\def\eps{\varepsilon}
\def\G{{\cal G}}

\def\P{{\mathsf{P}}} 
\def\S{{\cal S}}

\advance \hoffset 0.3in 
\advance \voffset 0.6in

\newtheorem{theorem}{Theorem}

\def\beqa{\begin{eqnarray*}}
\def\eeqa{\end{eqnarray*}}
\def\be{\begin{eqnarray}}
\def\ee{\end{eqnarray}}
\def\bg{\begin{gathered}}
\def\eg{\end{gathered}}

\begin{document}
\title{
Statistical Analysis of Link Scheduling \\ on Long Paths}

\author{
Yashar Ghiassi-Farrokhfal, ~J\"{o}rg Liebeherr, ~Almut Burchard
\vspace{-4mm}
\thanks{
Y. Ghiassi-Farrokhfal ({\tt yashar@comm.utoronto.ca}) and J. Liebeherr ({\tt jorg@comm.utoronto.ca}) are with the Department of Electrical and Computer Engineering, University of Toronto. A. Burchard ({\tt almut@math.utoronto.ca}) is with the Department of Mathematics, University of Toronto.
}
\thanks{
The research in this paper is supported in part by the 
Natural Sciences and Engineering Research Council of Canada (NSERC).}
}
\maketitle

\begin{abstract}
We study how the choice of packet scheduling algorithms 
influences end-to-end performance on long network paths. 
Taking a network calculus approach, we consider both 
deterministic and statistical performance metrics. 
A key enabling contribution for our analysis is a 
significantly sharpened method for computing 
a statistical bound for the service given to a flow by the network
as a whole.
For a suitably parsimonious traffic model we develop 
closed-form expressions for end-to-end delays, backlog, 
and output burstiness. The deterministic versions of our bounds 
yield optimal bounds on end-to-end backlog and output burstiness for 
some schedulers, and are highly accurate for end-to-end delay bounds. 
\end{abstract}

\section{Introduction}
\label{sec:intro}

Link scheduling algorithms for packet switches have been studied extensively, e.g., \cite{Zhang95}, 
however, little is known about the impact of link scheduling in large networks 
with long end-to-end paths. 
As a case in point, the packet dispersion of a CBR traffic flow at an overloaded  
First-in-First-Out (FIFO) link with cross traffic \cite{MeBjGu02} is 
given by 
\begin{align*}
r_{out} =     \frac{r_{in}}{r_{in}+r_c}C  \ ,  \notag 
\end{align*}
where $C$ is the constant-rate link capacity, 
$r_c$ is the arrival rate of CBR cross traffic,
and $r_{in}$ and $r_{out}$, respectively, are the 
arrival and output rate of the traffic flow (with $r_{in} + r_c> C$). 
Considering a path of a large number of overloaded FIFO links (with homogeneous 
cross traffic and link capacities), the output rate was shown in \cite{Ghiassi09} to 
converge to 
\begin{align*}
r_{out} \xrightarrow[H \rightarrow \infty]{}   \left[ C - r_{c} \right]_+ \ , \notag 
\end{align*}
where $H$ is the number of traversed links and $[x]_+ = \max \{x,0\}$.  
This is the same output rate observed in a network 
with priority scheduling 
where cross traffic is given higher priority. Thus, the question arises 
whether the role of link scheduling diminishes on long paths in more general 
settings. 

This paper develops an analysis that can assess 
the impact of a broad class of link scheduling algorithms on  end-to-end performance. 
Taking a network calculus approach \cite{Book-LeBoudec}, we characterize traffic 
in terms of {\em envelope} functions and service by  {\em service curve} functions.
We present bounds on end-to-end performance metrics for statistical, as well as 
worst-case assumptions on traffic.  
Our study considers all scheduling algorithms  for which the transmission order 
of backlogged packets is entirely determined by the difference of their arrival times.  
These schedulers, referred to as $\Delta$-schedulers, include FIFO, priority scheduling, and deadline based scheduling. 
A key contribution of this paper is a 
significantly sharpened method for computing 
a statistical bound on the service given to a flow on an entire network path from 
statistical characterizations of the service at each node of the path. 
For traffic models that are characterized by a long-term rate and a short-term burst, 
we derive closed-form 
expressions for statistical and worst-case bounds on end-to-end delay 
and backlog. 
We evaluate the tightness of the bounds by computing performance metrics 
for specific adversarial arrival scenarios. 
In some cases, e.g., for backlog bounds for some of the considered schedulers, 
we are able to show that our bounds are optimal. 

The analysis in this paper is made possible by two advances 
in the network calculus. The first is a tight description of schedulers by service 
curves at a single nodes from \cite{LiGhBu10}. The second is presented here.  
We advance the state-of-the-art of the stochastic network calculus \cite{Book-Jiang} 
by improving methods for concatenating statistical per-node service descriptions to achieve 
service descriptions for an entire network path. 
Compared to the best available methods \cite{CiBuLi06,LiGhBu10}, we achieve significantly improved end-to-end 
performance bounds for many scheduling algorithms. 
For the special case of deterministic 
scenarios (i.e., where bounds are never violated), we recover worst-case end-to-end delay bounds
known for FIFO networks \cite{Fidler03,Lenzini05}, while extending these bounds to 
more complex scheduling algorithms. 
For FIFO scheduling, researchers have found conditions when the 
output of a node has similar characteristics as the input, justifying a decomposition 
analysis where each node can be analyzed in isolation \cite{Roberts01,Eun05,Mazum05}. 
Our results indicate that the conditions 
for  decomposition are often met in non-FIFO networks, as long as the 
scheduling in the network is sufficiently distinct from a network with strict priorities.

Our study provides new insight into the role of scheduling 
algorithms in networks with long paths. For a network that is not 
permanently overloaded, we find that the differences between 
delay and backlog bounds at a single node 
persist in a multi-node setting. In fact, for the traffic and 
service models considered in this paper, the differences between bounds 
for various schedulers grow proportionally with the path lengths. 
 
The remaining sections of this paper are 
structured as follows. In
Sec.~\ref{sec:calculus}, we discuss the probabilistic 
traffic and service characterization in the network calculus, with traffic envelopes and 
service curves, and present our result of a new network service characterization. 
In Sec.~\ref{sec:delta}, we describe the class of $\Delta$-scheduling algorithms, 
and derive a network service description for a path of nodes with such schedulers. 
In Sec.~\ref{sec:bounds}, 
we derive closed-form end-to-end bounds on backlog, delay and the burstiness of 
output from a network.  In Sec.~\ref{sec:tight}, we derive lower bounds, thus 
enabling a discussion of the tightness of our results. 
We present numerical examples in Sec.~\ref{sec:eval} and conclude 
the paper in Sec.~\ref{sec:conclusions}.

\section{Network Calculus Framework}
\label{sec:calculus}
We take a network calculus modeling and 
analysis approach where arrivals and service of a flow are expressed 
in terms 
of deterministic or probabilistic bounds, referred to as 
{\em traffic envelopes} and {\em service curves}, respectively \cite{Book-LeBoudec}. 
We next discuss needed  definitions and concepts of 
the deterministic and stochastic network calculus. 
The last subsection 
contains the key  contribution of this paper: 
a new result to compute statistical service curves for a 
network path, that enables us to provide 
a sharpened analysis of end-to-end performance bounds.  

\subsection{Traffic Envelopes}
\label{sec:envelopes}

Consider a buffered link with a link scheduling algorithm, 
referred to as a node. Using 
a left-continuous time model, a sample path 
of the arrivals to the 
node in the time interval $[0,t)$ is denoted by $A(t$). 
We assume that traffic arrivals can be described by a stationary random
process, or at least satisfy stationary bounds.
Traffic departing from a node in $[0,t)$ is denoted
by $D(t)$. Both $A(t)$ and $D (t)$ are
nondecreasing functions with $A(0)=D(0)=0$, and we have $D (t)\leq A (t)$ for all $t\geq 0$. 
For brevity, we use the notation  $A(s,t)= A(t) - A(s)$ and $D(s,t)= D(t) - D(s)$ for arrivals 
and departures in the time interval $[s,t)$

We characterize arrivals to a node in terms of bounds on sample paths. 
A {\em deterministic sample path envelope}  $E$ provides a worst-case description of traffic in the sense that it satisfies for all $t\geq 0$
\begin{equation}
 \sup_{0\leq s\leq t}
\left\{A(s,t) - E (t - s) \right\} \leq 0 \ , 
\label{eq:detenv}
\end{equation}
with $E (t) = 0$ if $ t \leq 0$.
By definition, traffic arrivals never violate a deterministic envelope.

For a statistical network analysis we need a  probabilistic analogue,
i.e., a  bound for all sample paths of the arrivals which may be
violated with a small probability.
A {\em statistical sample path envelope}
$\G$ is a function that satisfies for all $ t \geq 0$
\begin{align}
\label{eq:sample-path}
 \P\Bigl(\sup_{0\leq s\leq t} \left\{ A(s, t) -\G(t - s; \sigma)\right\} 
> 0\Bigr)  \leq  \eps (\sigma) \ ,
\end{align}

\vspace{-1mm}
\noindent
where
$\sigma$ represents one or several parameters, and $\eps(\sigma) \geq 0$ is a bounding function. 
Further, $\G(t;\sigma)=0$ if $t\leq 0$. 
We can view $\eps(\sigma)$ as a bound on the probability of violating the envelope. 
The notation
$\G(t; \sigma)$ and $\eps(\sigma)$ indicate that both the  envelope and the bounding function depend on the same parameter(s). 
Setting $\eps(\sigma)=0$ recovers a deterministic sample path envelope. 

For numerical evaluations and optimizations, we will use  a parsimonious envelope characterization 
for traffic flows 
consisting of a rate parameter $\rho > 0$ and a burst parameter $\sigma \geq 0$, such that 
for all $s, t \geq 0$, 
\begin{align}
A (s, t)  \leq \rho (t -s ) + \sigma . 
\end{align}
A deterministic sample path bound for such an arrival characterization is given  by 
the envelope $E (t) = \rho t + \sigma$ (for $t>0$). 
This bound corresponds to the 
maximal output of a leaky bucket traffic regulator with rate $\rho$ and burst size $\sigma$. 

The probabilistic version of this bound takes the form that for all $s, t \geq 0$,  
\begin{align}
\label{eq:sample-path-sigma}
 \P\bigl(A(s, t) - \rho (t-s) > \sigma\bigr)  \leq  \eps (\sigma) \ , 
\end{align}
which can be read as the probability that the arrivals violate a long-term 
rate $\rho$ by more 
than $\sigma$ being bounded by $\eps (\sigma)$. 
When we assume that $\eps (\sigma)$ has an exponential decay, 
i.e., $\eps (\sigma)= Me^{-\alpha \sigma}$ for given constants $M$ and $\alpha$, we obtain the  
Exponential Bounded Burstiness (EBB) traffic model \cite{Yaron93}, which 
is related to the linear envelope characterization by Chang \cite{Chang94}.
We will write $A \sim (\rho,\alpha,M)$ to indicate that $A$ is an EBB flow with parameters $\rho$, $\alpha$, and 
$M$. 
The EBB model can express non-trivial processes, such as Markov-modulated On-Off processes \cite{Chang94}, 
but it does not apply to heavy-tailed or long-range correlated traffic.
Extensions of the EBB traffic model  have been proposed for 
bounding functions with faster than polynomial decay \cite{StaSi00} 
and even heavy-tailed decay \cite{LiBuCi10}. 

While a bound as in Eq.~\eqref{eq:sample-path-sigma} is not a 
statistical sample path envelope (satisfying Eq.~\eqref{eq:sample-path}), 
such an envelope can be constructed with an application of the union 
bound \cite{CiBuLi06}. 
For the EBB traffic model this yields a statistical sample path envelope 
\begin{align}
\G (t; \sigma)  = (\rho  + \gamma) t + \sigma \ , \quad
\eps (\sigma)  = M e \Bigl( 1 + \frac{\rho}{\gamma}\Bigr) e^{-\alpha \sigma} \ 
\label{eq:ebb-samplebound} 
\end{align}
for any choice of $\gamma >0$, where $\gamma$ can be viewed as a rate relaxation.

\subsection{Service Curves}
\label{sec:servicecurve}

We use the concept of a service curve \cite{Cruz95} to describe a lower bound on the service available to a flow at a node or sequence of nodes.
A node offers a {\em deterministic service curve} $\S$, if the input-output relationship of traffic is such that for all $t\geq 0$
\begin{align}
D(t) \geq \inf_{0\leq s \leq t} \left\{ A(s) + \S(t-s) \right\}  \ .
\label{eq:detserv}
\end{align}
The term on the right-hand side is referred to as min-plus convolution of 
$A$ and $\S$, and denoted by `$A \ast \S$'. Simple examples of service curves describe   
constant rate links, with $\S(t) = C t$,  and latency-rate servers,  
with $\S(t) = [ r (t -d)]_+$, for suitable non-negative constants $C, r,$ and $d$. 

A probabilistic extension of this concept leads to the formulation of a {\em statistical service curve} which satisfies for all $t \geq 0$ that
\begin{align}
\label{eq:srv-curve1}
\P\Bigl(D(t) < A * \S (t; \sigma)\Bigr) \leq  \eps(\sigma),
\end{align}
where the violation probability and the service curve may depend on one or more parameters indicated by  $\sigma$.

\subsection{Performance Bounds}
\label{subsec:bounds}

The network calculus seeks to provide upper bounds on the delay, backlog, and output 
burstiness for a flow at a node or network path. 
For  arrival and departure functions $A$ and $D$, the backlog $B(t)$ and 
the delay $W(t)$ at time $t$ are given by
\begin{align*}
& B(t)  = A(t) - D(t) \, ,  \\
& W(t)  = \inf \{ s \ge0\ | \ D(t+s) \geq A(t) \} \ . 
\end{align*} 
The backlog $B$ and the delay $W$ can be depicted, respectively, as the vertical and horizontal 
distance between the graphs of the arrival and departure functions. 
We use the term {\em output burstiness} to characterize a bound on the output process~$D$. 
Bounds on the above metrics can be formulated using the min-plus deconvolution operator  `$\deconv$', which for any  non-negative functions $f$ and $g$ is defined as $f \deconv g (t) = \sup_{s\geq 0} \{ f(t+s)-g(s)\}$.
For deterministic arrival and service 
characterizations, given by Eqs.~\eqref{eq:detenv} and~\eqref{eq:detserv}, 
such bounds can be found in \cite{Book-LeBoudec}. Below we provide the corresponding 
expressions for statistical arrival envelopes and service curves~\cite{CiBuLi06}. 
The service curve definition in \cite{CiBuLi06} is not identical to  
that in Eq.~\eqref{eq:srv-curve1}. At the same time, the proof of the bounds are close enough to be omitted here. 

Given a flow with  a statistical sample path envelope with $\G (t ; \sigma_g)$ and $\eps_g(\sigma_g)$, and a statistical  service curve with $\S (t ; \sigma_s)$ and $\eps_s(\sigma_s)$. 
Then with bounding function 
\[
\eps (\sigma_g, \sigma_s) = \eps (\sigma_g) + \eps (\sigma_s) \ , 
\]
the following bounds hold for all $s, t$ with $ 0 \leq s \leq t$. 
\begin{align*}
\intertext{$\bullet$  {\sc Output Burstiness:}}
& \P \bigl( D(s,t) > \G \deconv \S ( t- s; \sigma_g, \sigma_s ) \bigr) \leq \eps (\sigma_g, \sigma_s) \ ; \\ 
\intertext{ $\bullet$ {\sc Backlog: }}
& \P \bigl( B(t) > \G \deconv \S ( 0; \sigma_g, \sigma_s ) \bigr) \leq \eps (\sigma_g, \sigma_s) \ ; \\ 
\intertext{ $\bullet$ {\sc Delay: }} 
& \P \bigl( W(t) > d(\sigma_g, \sigma_s ) \bigr) 
\leq \eps (\sigma_g, \sigma_s) \ ;
\end{align*} 
where 
\[
d ( \sigma_g, \sigma_s ) = \min \left\{ s \, | \, \forall t \geq 0: \ \S (t+s;  \sigma_s) \geq 
\G (t ; \sigma_g) \right\} \ .
\] 
Setting $\eps (\sigma_g, \sigma_s)=0$ yields the deterministic worst-case bounds 
of the deterministic network calculus. 

\subsection{A New Convolution Theorem}
\label{sec:convolution}

The main advantage of working with service curves  is the ability to express the end-to-end service available 
on a network path as a min-plus convolution of the service available at each node 
of the path. 
In the deterministic network calculus, if each node on a path of $H$ nodes 
offers a service curve of $\S_h$ ($1 \leq h \leq H$), 
a service curve for the entire path is given 
by $\S_{\rm net} = \S_1 \conv \S_2 \conv \ldots \conv \S_H$. 
We refer to $\S_{\rm net}$ as {\em network service curve}.

Finding a network service curve in a statistical setting is 
difficult, and generally requires to introduce additional assumptions \cite{Book-Jiang}.  
A main result of this paper is a novel convolution theorem for a statistical 
network service curve, which will enable 
us to create improved statistical end-to-end bounds for many scheduling algorithms. 

\begin{theorem}[\bf Network Service Curve]  
\label{thm:convolve}
Consider a flow passing though a sequence
of $H$ nodes. Assume that at each node, the flow receives
a statistical service curve $\S_h(t;\sigma_h)$
with bounding function $\eps_h(\sigma_h)$, where $\sigma_h$ is a single non-negative 
parameter.
For each $h<H$, assume that
the bounding functions satisfy the tail estimate
\[ 
\int_0^\infty \eps_h(y)\, dy<\infty\,, 
\] 
and that the service curves $\S_h$ are non-increasing in $\sigma_h$ with
\begin{align}
\label{eq:S-property}
\S_h(t;\sigma_h+y)\ge \left[ \S_h(t;\sigma_h)-y\right]_+\,.
\end{align}
Let $\tau_1,\dots, \tau_{H-1}$ and $\gamma_1,\dots,\gamma_{H-1}$
be arbitrary positive parameters.
Then, for the entire path, the flow is guaranteed the statistical 
service curve 
\begin{align}\label{eq:conv-formula}
\S_{\rm net}(t;\sigma_1,\dots, \sigma_H) & = 
\Bigl[
\S_1 (\, . \,; \sigma_1)*\cdots * \S_H (\, . \,; \sigma_H) *\delta_{\sum_{h<H}\tau_h}(t)-\sum_{h<H}\gamma_h t\Bigr]_+\,, 
\end{align}
with bounding function
$$
\eps_{\rm net}(\sigma_1,\dots,\sigma_H)=\eps_H(\sigma_H)
+\sum_{h<H} \frac{1}{\gamma_h\tau_h}\int_{\sigma_h}^\infty \eps_h(y)\, dy\,.
$$
\end{theorem}
In the theorem, the function $\delta_a$ 
is defined as 
\begin{align}\label{eq-delta-function}
\delta_a(t) = 
\begin{cases}
0 \ , & t \leq a \ , \\
\infty \ , & t > a \ ,  
\end{cases}
\end{align}
for a constant $a \geq 0$.
The min-plus convolution of a function $f$ with $\delta_a$ corresponds to a 
right-shift of the function by $a$, that is, $f \conv \delta_a (t) = f (t -a)$.

In case of a deterministic bound, 
obtained by first setting $\eps_h(\sigma)=0$ for all $\sigma>\sigma_h$, and then taking 
$\tau_h=\gamma_h=0$, we recover the classical deterministic
convolution result for composing a network service curve.  

The theorem above generalizes a convolution theorem from~\cite{CiBuLi06},
by  allowing more freedom in the construction of service curves. 
In \cite{CiBuLi06}, service curves have the specific form 
$\S_h (t ; \sigma_h) = [\S_h (t) - \sigma_h]_+$, but \cite[Theorem~1]{CiBuLi06} 
performs the convolution of service curves only for the functions  $\S_h (t)$, without 
the $\sigma_h$'s. In contrast, the $\sigma_h$'s enter into the convolution of 
Eq.~\eqref{eq:conv-formula}. 
While carrying these parameters increases the complexity of the convolution,  
it yields  network service curves that can more closely describe the
behavior of certain schedulers. 

In applications, we will make specific choices for
$\tau_h$ and $\gamma_h$ that simplify the computation. 
We will also choose $\sigma_1,\dots, \sigma_H$
as explicit functions of a single parameter $\sigma$.
The impact of the  new convolution on end-to-end delay computations  
is evaluated in Sec.~\ref{sec:eval}. 

\begin{proof}
We first show without using
the assumption in Eq.~(\ref{eq:S-property}), that a network service 
curve is given by
\begin{align}
\label{eq:conv-proof-1}
S_{1\dots H}(t; \sigma_1, \dots, \sigma_H)
& = \inf_{x_1,\dots, x_H}\Bigl \{
S_H(x_H;\sigma_H)+ 
\sum_{h<H}S_h\Bigl(x_h; \sigma_h+\gamma_h\sum_{k>h}
(\tau_{k-1}+x_k)\Bigr) \Bigl\}\,,
\end{align}
where the infimum ranges over the set
$ \left\{x_1,\dots, x_H\ge 0 \ \Big\vert\ 
\sum_{h\le H} x_h + \sum_{h<H}\tau_h \le t\right\}$,
and the bounding function is as in the statement of the theorem.
The claim follows from there by observing that 
$\sum_{k>h}(\tau_{k-1}+x_k) \le t$ in the argument of $S_h$, 
and then using Eq.~(\ref{eq:S-property}).

To establish Eq.~(\ref{eq:conv-proof-1}),
we proceed by induction.  
For $H=1$, there is nothing to show.
For $H=2$ nodes, we need to show that
$$
S_{12}(t; \sigma_{1}, \sigma_2)
= \inf_{x_1+x_2\le t-\tau_1}\left \{
S_2(x_2;\sigma_2)+ S_1\Bigl(x_1; \sigma_1+\gamma_1 (\tau_1+x_2)\Bigr) 
\right\}
$$
satisfies
\begin{align}
\label{eq:conv-proof-2}
P\{D_2(t)<A_1 * S_{12}(t;\sigma_{1}, \sigma_2) \}& \le \eps_2 (\sigma_2) + \frac{1}{\gamma_1\tau_1}
\int_{\sigma_1}^\infty \eps_1(y)\, dy + \eps_2(\sigma_2) 
=: 
\eps_{12}(\sigma_1,\sigma_2)\,.
\end{align}
To see this, note that for every sample path 
where $D_2(t)\ge A_2*S_2(t;\sigma_2)$, there exists
exists a point $x_2\le t$ such that
\begin{equation}
\label{eq:conv-proof-3}
D_2(t)\ge A_2(t-x_2)+S_2(x_2;\sigma_2)\,.
\end{equation}
If, additionally, 
\begin{align}
\label{eq:conv-proof-4}
& \forall x\le t\ \  \exists x_1\le x:\quad 
D_1(t-x)\ge A_1(t-x_1-x-\tau_1) + S_1(x_1; \sigma_1+\gamma_1(x+\tau_1))\,,
\end{align}
then, since $A_2=D_1$, it follows that
\begin{equation} \label{eq:conv-samplapath}
D_2(t)\ge A_1*S_{12}(t; \sigma_1, \sigma_2)\,.
\end{equation}
We claim that the event in Eq.~(\ref{eq:conv-samplapath})
occurs at least with
probability $1-\eps_{12}(\sigma_1,\sigma_2)$. 
By definition, the probability that the event in 
Eq.~(\ref{eq:conv-proof-3}) fails is bounded by
$$
P(D_2(t)< A_2*S_2(t;\sigma_2))\le \eps_2(\sigma_2)\,.
$$
To estimate the event in Eq. (\ref{eq:conv-proof-4}),
we discretize the time variable $x$ by choosing an integer
$k\ge 1$ such that
$(k-1)\tau_1 \le x < k \tau_1$. If Eq. (\ref{eq:conv-proof-4})
fails, then the monotonicity of $D_1$,  $A_1$, and $S_1$
implies that there exists a $k\ge 1$ such that
$$
D_1(t-k\tau_1) <A_1(t-x_1-k\tau_1) + S_1(x_1; \sigma_1+\gamma\tau_1)\,.
$$
We conclude with the union bound that Eq.~(\ref{eq:conv-proof-4})
fails with probability at most
\begin{align*}
&\hspace {-1cm}
\sum_{k=0}^\infty P\Bigl(
D_1(t-k\tau_1) < \inf_{x_1\le t-k\tau_1}
\left\{
A_1(t-x_1-k\tau_1) 
+ S_1(x_1; \sigma_1+ k \gamma_1 \tau_1)\right\}\Bigr)\\
& \le \sum_{k=1}^\infty 
\eps_1(\sigma_1+k \gamma_1 \tau_1)\\
&\le \frac{1}{\gamma_1\tau_1} \int_{\sigma_1}^\infty \eps_1(y)\, dy\,.
\end{align*}
Eq.~(\ref{eq:conv-proof-2}) follows with another 
application  of the union bound.

For the inductive step, let $H>2$,
and assume that Eq.~(\ref{eq:conv-proof-1}) holds 
for every sequence of $H-1$ nodes. Given
 $H$ nodes, we first apply the inductive hypothesis to obtain
for the last $H-1$ nodes the service curve
\begin{align*}
 S_{2\dots H}(t; \sigma_2, \dots, \sigma_H) & =
\inf_{x_2,\dots, x_H}\biggl\{
S_H(x_H;\sigma_H) 
+ \sum_{1<h<H}S_h\Bigl(x_h; \sigma_h+\gamma_h\sum_{k>h}
(\tau_{k-1}+x_k)\Bigr) \biggr\}\,,
\end{align*}
where the infimum ranges over 
$\left\{x_2,\dots, x_H\ge 0 \ \Big\vert\ 
\sum_{h\le H} x_h + \sum_{h<H}\tau_h \le t\right\}$,
and the bounding function is
$$
\eps_{2\dots H}(\sigma_2,\dots, \sigma_H)
= \eps_H(\sigma_H)+\sum_{1<h<H} \frac{1}{\gamma_h\tau_h}
\int_{\sigma_h}^\infty \eps_h(y)\, dy\,,
$$
and then use the $H=2$ case with
$S_{2\dots H}$ and $\eps_{2\dots H}$
in place of $S_2$ and $\eps_{2}$.
\end{proof}

\begin{figure}[thb]
\centering
  \includegraphics[width=0.5\columnwidth]{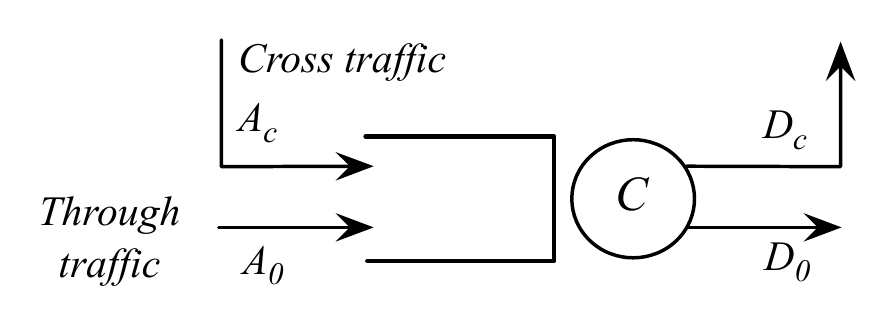}
  \caption{Buffered link with fixed capacity $C$ and cross traffic.}  
\label{fig:experiment_setup_1hop}
\end{figure}

\section{Networks of $\Delta$-Schedulers}
\label{sec:delta}

In this section, we obtain a statistical network service curve 
that describes the service available on an end-to-end path of 
network links, where each link realizes a packet scheduling algorithm. 
To characterize link schedulers, we use a general description of scheduling 
algorithms, referred to as  $\Delta$-scheduling, recently developed in \cite{LiGhBu10}. 

\subsection{$\Delta$-Schedulers}

Consider a buffered link with a fixed capacity of rate $C$  
and without buffer size constraints as shown in 
Fig.~\ref{fig:experiment_setup_1hop}. 
We are interested 
in describing the performance of a (through) flow with arrival function $A_0$ at this link. 
The flow experiences cross traffic from a flow with arrival function $A_c$. 
We do not assume independence of through and cross traffic. On the contrary, the 
traffic can be correlated in an adversarial fashion. 
A scheduling algorithm at the link determines 
the order of transmission of backlogged traffic. 

We consider the class of work-conserving scheduling algorithms whose operation can be entirely 
described by a set of constants $\Delta_{ij}$ where $i$ and $j$ indicate  flow indices. 
For two backlogged packets  $p_i$ and $p_j$ from flows $i$ and $j$,  
the scheduler selects $p_i$ over $p_j$ for transmission, if 
$p_j$ arrives more than $\Delta_{ij}$ time units after $p_i$. 
Schedulers that can be described in this fashion are referred to as {\em $\Delta$-schedulers}. Special 
cases are FIFO ($\Delta_{ij}=0, \forall i,j$) and  
Static Priority ($\Delta_{ij}=\infty$, if the priority of flow~$i$ is lower than that of flow $j$, 
and $\Delta_{ij}=-\infty$, if the priority of flow~$i$ is higher than that of flow~$j$). 
A $\Delta$-scheduler with non-zero but finite values of $\Delta_{ij}$  can 
be implemented by an algorithm that tags each packet arrival
with a timestamp consisting of the sum of the arrival time and a flow-specific 
value (e.g., a target  delay), and transmits packets in the order of 
timestamps. Here, the value of $\Delta_{ij}$ is the 
difference of the tags assigned to through and cross traffic. 
For example, the Earliest-Deadline-First 
can be realized by setting $\Delta_{ij}=d^*_i - d^*_j$, where 
$d^*_i$ and $d^*_j$ are the a priori delay bounds of 
flows $i$ and $j$. 

For any  $\Delta$-scheduler operating at a buffered link with capacity $C$ 
where the cross traffic 
is bounded by a statistical sample path envelope given by $\G_c(t;\sigma)$ and 
$\eps_c(\sigma)$, 
it was shown in~\cite{LiGhBu10} that, for any 
choice of $\theta \geq 0$, the function
\be
\label{eq-leftover-srvcrv} 
\S (t;\sigma) = \Bigl[Ct - 
\G_c(t - \theta +\Delta(\theta); \sigma) \Bigr]_+
I_{t>\theta} \ ,  
\ee 
is a statistical service curve with bounding function $\eps_c (\sigma)$, 
and $I_{\rm expr}$ denotes the indicator function, 
with $I_{\rm expr}=1$ if the expression `${\rm expr}$' is true, 
and $I_{\rm expr}=0$ otherwise. 
Also, $\Delta(\theta) = \min \{\theta, \Delta \}$, 
where $\Delta$ is the constant for the precedence of the through traffic with arrival function $A_0$ 
relative to the cross traffic with arrival function $A_c$. (Note: We denote $\Delta = \Delta_{0c}$ since 
we only work with the constant for arrivals $A_0$ relative to $A_c$.) 
For a single node, the service curve characterization from Eq.~\eqref{eq-leftover-srvcrv} 
was shown to 
yield a tight delay analysis, provided that cross traffic and through traffic are  characterized by concave 
deterministic sample path envelopes \cite{LiGhBu10}. 

\begin{figure}[t]
\centering
  \includegraphics[width=0.8\columnwidth]{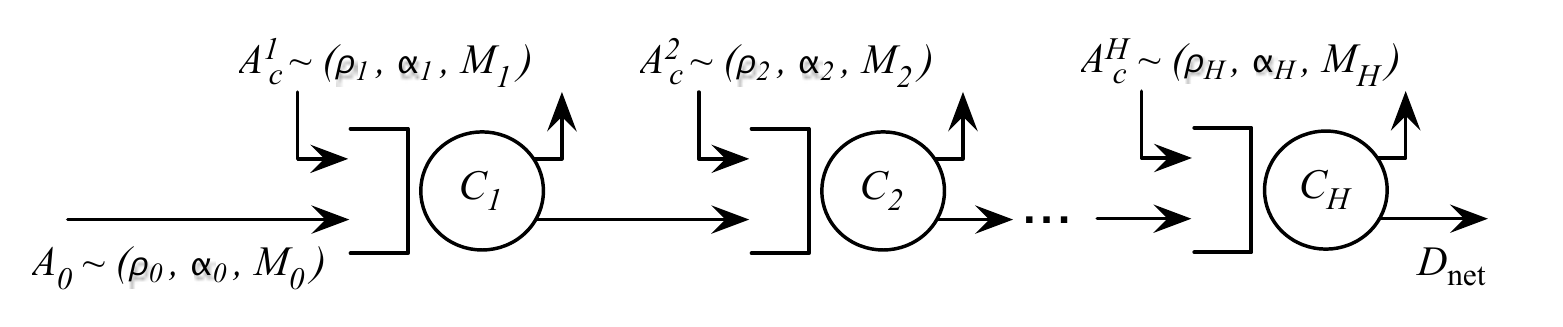}
  \caption{Network path with cross traffic at each link.}  
\label{fig:case-study}
\end{figure}
\subsection{A Network Service Curve for $\Delta$-Schedulers}
\label{sec:e2e-analysis}

We next apply Theorem~1 to construct a network service curve for
$\Delta$-schedulers along a network path  as shown in 
Fig.~\ref{fig:case-study}, where the through flow traverses 
a tandem of $H$ buffered links with $\Delta$-scheduling, 
and with cross traffic. 
The arrivals of the cross flow and the 
$\Delta$-scheduler can be different at each link. 
We use $\Delta^h$ to denote the constant of the $\Delta$-scheduler at the $h$-th node. 
Arrivals  to the cross flow at the $h$-th node, $A^h_c$, conform
to an EBB model, as given by Eq.~(\ref{eq:sample-path-sigma}),
with rate $\rho_h$ and
exponentially decaying violation probability 
$\eps_h(\sigma)=M_h e^{-\alpha_h\sigma}$. 

Inserting the statistical sample-path envelope
for an EBB cross flow from Eq.~\eqref{eq:ebb-samplebound}   
into Eq.~(\ref{eq-leftover-srvcrv}), 
we obtain for the through flow at node~$h$ the service curve
and the bounding function 
\begin{align*}
\S_h(t; \sigma_h) & = 
\Bigl[C_h t- \Bigl((\rho_h\!+\!\gamma)
(t\!-\!\theta_h\!+\!\Delta^h(\theta_h))  
+\sigma_h\Bigr)I_{t>\theta_h\!-\!\Delta^h}\Bigr]_+I_{t>\theta_h} , \\ 
\eps_h(\sigma_h) 
& =M_h e\Bigl(1+\frac{\rho_h}{\gamma}\Bigr)
    e^{-\alpha_h\sigma_h}\,.
\end{align*}
Here, $\gamma >0$ is an arbitrary constant. 
To simplify the computation of the convolution,
we replace this service curve with the lower bound 
\begin{align}\label{eq-per-node-srv-crv-appx}
\S_h(t;\sigma_h)& = 
\Bigl[C_h t-[(\rho_h+\gamma)
(t-\theta_h+\Delta^h(\theta_h))+\sigma_h]_+\Bigr]_+I_{t>\theta_h}.
\end{align}
The replacement comes without loss for $\Delta^h \geq 0$, and 
a reduction by at most $(C_h-\rho_h)[\Delta^h]_-$
for $\Delta^h < 0$, 
where we use $[x]_- = \max \{-x, 0\}$. The key point is that the right hand side
of Eq.~(\ref{eq-per-node-srv-crv-appx})
is concave and strictly increasing
in $t$ as soon as it is positive.
We apply Theorem~\ref{thm:convolve} with the
free parameters set to
$\gamma_h=\gamma$ 
and $\tau_h= \frac{\alpha_h^{-1}}{\sum_{k<H} \alpha_k^{-1}} \tau_{\rm net}$
for $h=1,\dots, H-1$, where $\tau_{\rm net}$ and $\gamma$ are chosen below.
By Eq.~(\ref{eq:conv-formula}),
\begin{align}
\notag
\S_{\rm net}(t;\sigma) & =
\left[
\S_1*\cdots * \S_H *\delta_{\tau_{\rm net}}(t)-(H-1)\gamma t\right]_+\\
  & \ge \left[
\S_1*
\cdots * \S_H - (H-1)\gamma t\right]_+
*\delta_{\tau_{\rm net}}(t)
-(H\!-\!1)\gamma\tau_{\rm net}\,,
\label{eq:Snet-delta}
\end{align}
where we use $\sigma$ to mean $\sigma = (\sigma_1,\ldots,\sigma_H)$. 
To evaluate the term inside the square brackets
in the second line, we may equivalently
convolve the functions $[\S_h(t; \sigma)-(H-1)\gamma t]_+$.
For each of these factors, we write
\begin{equation}
\label{eq:S-h-net1}
\left[ \S_h(t;\sigma_h)-(H-1)\gamma t\right]_+ 
=\left[ (C_h-(H\!-\!1)\gamma)t  
-  [(\rho_h\!+\!\gamma)(t\!-\!\theta_h\!+\!\Delta^h(\theta_h))
+\sigma_h]_+\right]_+I_{t>\theta_h}  \,.
\end{equation}
If we choose $\theta_h$ large enough so that
\begin{align}
\underbrace{(C_h-(H\!-\!1)\gamma)\theta_h-
[(\rho_h\!+\!\gamma)\Delta^h(\theta_h)
+\sigma_h]_+ }_{=:U_h}\ge 0\,,
\label{eq:formula-Uh} 
\end{align}
then we can re-write Eq.~\eqref{eq:S-h-net1} as 
\[
\left[ \S_h(t;\sigma_h)-(H-1)\gamma t\right]_+ 
= \widetilde \S_h * \delta_{\theta_h}(t)\, , 
\]
with 
\begin{align}
\widetilde \S_h(t;\sigma_h)
& = \min\left \{(C_h-(H\!-\!1)\gamma)(t+\theta_h),
(C_h\!-\!\rho_h\!-\!H\gamma)t +U_h\right\}I_{t>0}\,.
\label{eq:S-h-tilde}
\end{align}
Since each of the $\widetilde \S_h$ is concave and strictly increasing in~$t$, 
 their convolution is given by the pointwise minimum 
\begin{align*}
\widetilde{\S}_{\rm net}(t;\sigma) &:=
\widetilde{\S}_1\conv \ldots  \conv \widetilde{\S}_H(t;\sigma)
 = \min_{h=1,\ldots,H}\Bigl\{ \widetilde{\S}_h (t;\sigma_h)\Bigr\} \,.
\end{align*}
Using the associativity and commutativity of the
min-plus convolution, we
obtain from Eq.~\eqref{eq:Snet-delta} the network service curve
\begin{equation}
\label{eq:Snet-tildeSnet}
\S_{\rm net}(t;\sigma)\ge  \bigl( \widetilde{S}_{\rm net}
\conv\delta_{\tau_{\rm net}+\sum_{h=1}^{H}\theta_h}\bigr) (t) - (H-1)\gamma \tau_{\rm net}\,.
\end{equation}
The bounding function is computed with Theorem~\ref{thm:convolve} as 
\begin{align*}
\eps_{\rm net}(\sigma)& =
M_H e \Bigl(1+\frac{\rho_H}{\gamma}\Bigr)e^{-\alpha_H\sigma_H}
+ \frac{\sum_{h<H}\alpha_h^{-1}}{\gamma  \tau_{\rm net}} \sum_{h=1}^{H-1} 
M_h e \Bigl(1+\frac{\rho_h}{\gamma} \Bigr)e^{-\alpha_h\sigma_h}\,.
\end{align*}

\noindent
A probabilistic network service curve 
for a cascade of $\Delta$-schedulers has been 
computed in \cite{LiGhBu10} by 
applying a convolution theorem for 
service curves from \cite{CiBuLi06}.
This required to bring the service curves into the form 
$\S_h (t ; \sigma_h) = [\S_h (t) - \sigma_h]_+$, 
which resulted in a 
pessimistic service curve of the form 
\begin{align*}
\S_h(t;\sigma_h)& = \left[C_h t-(\rho_h\!+ \!\gamma)
[t-\theta_h+\Delta^h(\theta_h)]_+ - \sigma_h\right]_+\!I_{t>\theta_h} \!. 
\end{align*}
While the difference in the placement of $\sigma_h$
compared to 
Eq.~\eqref{eq-per-node-srv-crv-appx} may appear subtle, the evaluation 
in Sec.~\ref{sec:eval} will show that the 
service curve together with the new convolution 
theorem (Theorem~\ref{thm:convolve}) results 
in substantially improved bounds.

\section{Output, backlog, and delay bounds}
\label{sec:bounds}

In this section, we apply the performance bounds (from 
Sec.~\ref{subsec:bounds}) with the network service curve for $\Delta$-schedulers 
(from Sec.~\ref{sec:e2e-analysis}) 
to a through flow that traverses  a  network as shown in Fig.~\ref{fig:case-study}. 
We will make  choices for the free parameters~$\theta_h$
appearing in Eq.~(\ref{eq-leftover-srvcrv}), so that we can obtain 
explicit bounds on the distribution of the
output burstiness from the last node, the end-to-end backlog, and the 
end-to-end delay.  The bounds apply to both deterministic 
(worst-case) and statistical assumptions. 

Consider the same network as described in Sec.~\ref{sec:delta}.
The through traffic is an EBB flow with arrival function $A_0 \sim (\rho_0, \alpha_0, M_0)$.
We assume that the stability condition
\[
\rho_0 <\min_{h=1,\ldots ,H} \{C_h-\rho_h\}\,
\]
holds, and choose $\gamma$ such that 
$$0< H\gamma< \min_{h=1,\ldots,H} \{ C_h-\rho_h\} -\rho_0  \,.
$$

To formulate the performance bounds, we need to
introduce some more notation. Let
\begin{align}
\alpha_{\rm net}&=\Bigl(\sum_{h=0}^H \alpha_h^{-1}\Bigr)^{-1}  \!\!,
C_{\rm net}\!=\!\min_{h=1,\ldots,H} \{C_h\}, \, \tau_{\rm net}\!=\!\frac{1}{\alpha_{\rm net}C_{\rm net}}\,, \notag \\
M_{\rm net}& = M_0e\Bigl(1+\frac{\rho_0}{\gamma}\Bigr)
+ M_He\Bigl(1+\frac{\rho_H}{\gamma} \Bigr) 
+ 
\frac{C_{\rm net}}{\gamma} \sum_{h=1}^{H-1} M_h e 
\Bigl(1+\frac{\rho_h}{\gamma}\Bigr)\,,\label{eq:Mnet}
\end{align}

\vspace{-3mm}\noindent
and choose

\vspace{-3mm}
\begin{align} 
\label{eq:sigma}
& \sigma_h=\frac{\alpha_{\rm net}}{\alpha_h} \sigma\,,\quad h=0,\dots, H\,. \notag 
\end{align} 

For any given value of $\sigma$,  the smallest value
of $\theta_h$
that makes $\widetilde \S_h(t)>0$ for all $t>0$ (see Eq.~\eqref{eq:S-h-tilde}) is given by
\begin{equation}
\label{eq:choose-theta-simple}
\theta^*_h =\min\left\{\frac{\sigma_h}{C_h-\rho_h -H\gamma},
\frac{[\sigma_h+(\rho_h +\gamma)\Delta^h]_+}{C_h-(H-1)\gamma}\right\}\,,
\end{equation}
and the corresponding value for $U_h$ is given by
\begin{equation} \label{eq:def-U-simple}
U^*_h =[\sigma_h+(\rho_h+\gamma)\Delta^h]_-\,. 
\end{equation}

For the network in Fig.~\ref{fig:case-study}, 
let  $D_{\rm net}$ denote the output of the through traffic 
from the last ($H$-th) node. 
Let $B_{\rm net}(t) = D_{\rm net} (t)-A_0(t)$ and 
$W_{\rm net}(t)$, respectively,  denote the total  
backlog and delay of the through traffic in the network of $H$ nodes.

\begin{theorem}[\bf Closed-form Bounds] 
\label{thm:closed-form} 
With the above notation and definitions, the following bounds hold.

\smallskip
\noindent $\bullet$ 
{\sc Output Burstiness:}
$$
\P\bigl(D_{\rm net}(s,t)-(\rho_0+\gamma)(t-s)>b_{\rm net}(\sigma)\bigr)
\le M_{\rm net}e^{-\alpha_{\rm net}\sigma}
\,,
$$
where

\vspace{-5mm}
\begin{align*}
b_{\rm net}(\sigma)&= (\rho_0+H\gamma)\tau_{\rm net}+\sigma_0
+ (\rho_0+\gamma) \sum_{h=1}^H \theta^*_h\,;
\end{align*}

\noindent $\bullet$ 
{\sc End-to-end backlog:}
$$
\P\bigl(B_{\rm net}(t) >b_{\rm net}(\sigma)\bigr)
\le M_{\rm net}e^{-\alpha_{\rm net}\sigma}\,;
$$

\noindent $\bullet$ 
{\sc End-to-end delay:}
$$\P\bigl( W_{\rm net}(t)>d_{\rm net}(\sigma)\bigr)
\le M_{\rm net}e^{-\alpha_{\rm net}\sigma}\,,
$$
where
\begin{align*}
d_{\rm net}(\sigma)&= \tau_{\rm net}+\max_{h=1,\dots, H}
\Bigr\{ \max\Bigl\{\frac{\sigma_0+ (H-1)\gamma\tau_{\rm net}}
{C_h-(H-1)\gamma},
\frac{\sigma_0+(H-1)\gamma\tau_{\rm net}-U^*_h}
{C_h-\rho_h -H\gamma}\Bigr\}
\Bigr\}+
\sum_{h=1}^H \theta^*_h\,.
\end{align*}
\end{theorem}

The output bound shows that
the departing traffic satisfies the EBB model.  
In the expressions for $b_{\rm net}(\sigma)$ and $d_{\rm net}(\sigma)$, 
the second term quantifies the contribution of the
through flow's burstiness to the delay at the bottleneck link,
and the last term sums up the contributions of the burstiness
of the cross flows. On long paths, i.e., for $H$ large, the last term  
will dominate the second term. 

An important observation is that 
all bounds increase with~$\Delta^h$, 
and that  the difference between
end-to-end delay and backlog bounds for different schedulers grows
linearly with the number of nodes. This clearly indicates that 
the impact of the scheduling algorithm on end-to-end performance
bounds does not diminish on long network paths for these schedulers. 

In a discrete-time setting (where $t = 0, 1, \ldots$), 
the bounds in the theorem hold with
$\tau_{\rm net}=0$, and with Eq.~(\ref{eq:Mnet}) replaced by
\begin{align*}
M_{\rm net}& = \frac{M_0}{1-e^{-\alpha_0\gamma}} 
+ \frac{M_H}{1-e^{-\alpha_H\gamma}} + 
\sum_{h=1}^{H-1} \frac{M_h}{(1-e^{-\alpha_h\gamma})^2}\,.
\end{align*}
\noindent 
\begin{proof} We insert 
the network service curve
from Theorem~1 into the 
performance bounds from Sec.~\ref{subsec:bounds}.
Since 
$\alpha_h\tau_h\leq 1 / C_{\rm net}$, the bounding  function 
evaluates to
$
\eps_0(\sigma_0)+ 
\eps_{\rm net}(\sigma)= M_{\rm net} e^{-\alpha_{\rm net}\sigma}\,.
$

The backlog bound is given by
$b_{\rm net}(\sigma) = \G_0\deconv S_{\rm net}(0)$.
For $\theta_h\ge \theta^*_h$, 
we compute this deconvolution as
\begin{align}
\notag b_{\rm net}(\sigma) & = \G_0 \deconv 
\left[ \bigl( \widetilde{S}_{\rm net}
\conv\delta_{\tau_{\rm net} +\sum_{h=1}^{H}\theta_h} \bigr)  -(H\!-\!1)\gamma\tau_{\rm net}\right]_+\!(0)\\
\label{eq:bsigma}
& \le  \G_0 \Bigl(\tau_{\rm net} +\sum_{h=1}^{H}\theta_h\Bigr) 
+(H-1)\gamma \tau_{\rm net}\,.
\end{align}
In the first line, we have
used our representation for the network service curve from Eq.~\eqref{eq:Snet-tildeSnet}. 
The key step is the second line, where we use
that $\widetilde{\S}_{\rm net}$ is concave
and $\G_0$ is convex for $t>0$, and that
$\G_0(t)<\S_{\rm net}(t)$ for $t$ is sufficiently large by
the stability condition and our choice of $\gamma$.
Plugging in $\theta_h=\theta^*_h$, we obtain the
desired backlog bound.
The same computation gives the bound on the output burstiness. 

We turn to the delay bound.
For $\theta_h\ge\theta^*_h$, the delay bound from 
Theorem~\ref{thm:closed-form} is given by 
$$
d_{\rm net}(\sigma)=\min\{d\ge 0 \mid \forall t\ge 0:
\S_{\rm net}(t+d;\sigma)\ge \G_0(t;\sigma_0)\}\,.
$$
By Eq.~(\ref{eq:Snet-tildeSnet}), and using as above the concavity
of $\widetilde{\S}_{\rm net}$ and the stability assumption, 
we arrive  at 
\begin{align}
\label{eq:thm1-delay-proof}
d_{\rm net}(\sigma)&=
\tau_{\rm net} + \sum_{h=1}^H \theta_h 
+ \min \Bigl\{ X\ge 0\mid \widetilde{\S}_{\rm net}(X)
\ge  \sigma_0 + (H-1)\gamma \tau_{\rm net} \Bigr\} \,.
\end{align}
Expressing the function $\widetilde{\S}_{\rm net}$ 
in terms of the $\widetilde{\S}^h$, we see that
the minimum is assumed for
\begin{align*}
X&=\max_{h=1,\dots, H} \Bigl\{ \max\Bigl\{
\frac{ \sigma_0 + (H-1)\gamma\tau_{\rm net}} {C_h-(H-1)\gamma} -\theta_h, 
\frac{\sigma_0 + (H-1)\gamma\tau_{\rm net}-U_h}{C_h-\rho_h-H\gamma}\Bigr\}\Bigr\}\,.
\end{align*}
Inserting $\theta_h=\theta^*_h$ and $U_h=U^*_h$
for $h=1,\dots, H$ completes the proof.
\end{proof}

\bigskip
We note that $\theta_h=\theta^*_h$
is the optimal choice for the backlog and output bounds.
To see this, observe that a selection in the range  
$0 \leq \theta_h<\theta^*_h$ decreases 
$\widetilde \S_h$, resulting in a smaller network service curve.  On the other hand, from Eq.~(\ref{eq:bsigma}) it is apparent that $b_{\rm net}(\sigma)$ increases for $\theta_h>\theta^*_h$.
We will see in the next section that the backlog
bound is highly accurate, and in many cases tight. 

For the delay bound in Eq.~\eqref{eq:thm1-delay-proof}, the optimal choice for~$\theta_h$ is not 
obvious, since~$U_h$ can be increased by taking~$\theta_h>\theta^*_h$, which may result 
in a smaller value for $d_{\rm net}(\sigma)$.
The choice $\theta_h=\theta^*_h$ is
clearly suboptimal when $C_h-\rho_h$ is small and the path is short,
i.e., when the cross traffic consumes most of the bandwidth.
In the case of a single node ($H=1$) and 
deterministic arrival bounds,
choosing $\theta_1=d_{\rm net}$ recovers known necessary and
sufficient conditions for delay bounds  
e.g., \cite{Cruz91}. For long paths ($H>1$ large),
it is plausible that the choice 
$\theta_h=\theta^*_h$, which is optimal for
backlog and output, also leads to acceptable delay bounds.
We will see in the next section that this is indeed the case.

Stronger delay bounds than those stated in the theorem can be obtained
by solving the following optimization problem:
\begin{align}
\mbox{minimize} \quad  d_{\rm net}(\sigma) =  \tau_{\rm net}+ 
X+\sum_{h=1}^H \theta_h \ , \label{eq-opt-prob-1}
\end{align}
subject to the constraints that $X\ge 0$ and

\medskip
\noindent 
\begin{align}
\forall h: 
\begin{cases} & (C_h-(H\!-\!1)\gamma)(X+\theta_h) \geq \sigma_0+ (H-1)\gamma\tau_{\rm net}\,, \nonumber \\ 
& (C_h-\rho_h-H\gamma)X+U_h\geq \sigma_0 +
(H-1)\gamma\tau_{\rm net}
\,,\nonumber \\ 
 & \theta_h \geq \theta_h^*\,, 
 \end{cases} \nonumber
 \end{align}

\smallskip
\noindent 
where $U_h$ and $\theta_h^*$ 
are defined in Eqs.~\eqref{eq:formula-Uh} and~\eqref{eq:choose-theta-simple}.  
This optimization problem is convex if all $\Delta^h$ are non-positive, 
but convexity does not hold if $\Delta^h > 0$ for 
at least one node. Nevertheless, the problem
is tractable, because the objective function is linear
in the variables $X$ and $\theta_1,\dots, \theta_H$.
We will see in the evaluation that this optimization 
sometimes (i.e., for $H$ small) improves the closed-form 
delay bound of Theorem~\ref{thm:closed-form}. 
For FIFO networks (all $\Delta^h=0$) and deterministic 
traffic  ($\gamma=\tau_{\rm net}=0$) the 
optimization recovers existing deterministic end-to-end delay bound for feedforward 
networks in  \cite{Fidler03,Lenzini05}. 

\medskip
\section{Tightness of End-to-end Bounds }
\label{sec:tight}

We now address the tightness of the
bounds in Theorem~\ref{thm:closed-form}. We consider  the 
same network as in the previous section, but assume a 
 deterministic context,
where the arrivals of the through traffic and cross traffic 
at the $h$-th node satisfy the deterministic sample path 
envelopes $E_0 (t)= \sigma_0+\rho_0t$ and 
$E_h(t)=\sigma_h+\rho_ht$ ($t >0$), respectively.
We denote a deterministic sample path envelope for 
the departing through traffic at the $H$-th node by $E_{\rm net}(t)$. 
We obtain worst-case bounds
by setting $\gamma =\tau_{\rm net}=0$ in the 
expressions for $\theta^*_h$ in Eq.~\eqref{eq:choose-theta-simple} 
and $U^*_h$ in Eq.~\eqref{eq:def-U-simple}, yielding 
$$
\theta^*_h =\min\left\{\frac{\sigma_h}{C_h-\rho_h}, 
\frac{[\sigma_h+\rho_h \Delta^h]_+}{C_h}\right\}\,,
$$
and
$$
U^*_h= [\sigma_h+\rho_h\Delta^h]_-\,.
$$
Also, setting $\gamma =\tau_{\rm net}=0$ in 
Theorem~\ref{thm:closed-form}, 
we obtain the following bounds, denoted by $E^*_{\rm net}, b^*_{\rm net}$, 
and $d^*_{\rm net}$.

\medskip
\begin{itemize}

\item 
{\sc Output Burstiness:}
\[
E^*_{\rm net}(t) = \sigma_0 + \rho_0 \sum_{h=1}^H 
\theta^*_h + \rho_0 t\,;
\]

\item {\sc End-to-end backlog:}
\begin{equation}
b^*_{\rm net} = \sigma_0 + \rho_0 \sum_{h=1}^H \theta^*_h\,;
\label{eq:Bnet-simple}
\end{equation}

\item {\sc End-to-end delay:}
\begin{align} 
 \hspace{-5mm} d^*_{\rm net} = \max_{h=1,\dots, H}
\Bigl\{ \max\Bigl\{\frac{\sigma_0}{C_h}, \frac{\sigma_0-U^*_h}{C_h-\rho_h }\Bigr\} \Bigr\} 
+ \sum_{h=1}^H \theta^*_h\,.
\label{eq:Wnet-simple}
\end{align}
\end{itemize}

For the special cases of priority scheduling where 
through traffic has lower priority ($\Delta^h=+\infty$) or higher
priority ($\Delta^h=-\infty$), we obtain 
$$
W_{\rm net}(t)\le
\begin{cases}
\displaystyle{\frac{\sigma_0}{\min\{C_h-\rho_h\}}
+ \sum_{h=1}^H \frac{\sigma_h}{C_h\!-\!\rho_h }\,,}
& \Delta^h=+\infty\,, \\
\displaystyle{
\frac{\sigma_0}{\min\{C_h\}} \,,}
&\Delta^h=-\infty \,.
\end{cases}
$$
These bounds are known to be tight, which  
can be verified with elementary methods of the deterministic network calculus~\cite{Book-LeBoudec}. 

\medskip 
In the remainder of this section, we evaluate the tightness of the bounds 
for $\Delta$-schedulers with finite values of $\Delta$. 
For this, we consider specific adversarial sample path scenarios for through 
and cross traffic, and compute the maximal observed end-to-end delay and backlog. 
For any specific arrival scenario, let $B_{\rm max}=\max_t\{B_{\rm net}(t)\}$
and $W_{\rm max}=\max_t\{W_{\rm net}(t)\}$ denote the maximal 
end-to-end backlog and delay, respectively, of the through traffic. 
Clearly, any $B_{\rm max}$ and $W_{\rm max}$, respectively,  are lower bounds for the 
worst-case end-to-end backlog and delay. 

\begin{theorem}[\bf Tightness of Bounds] For 
$-\infty < \Delta^h < \infty$, there exists a scenario
of through and cross traffic that satisfies the assumptions
stated above, such that
\begin{align} 
B_{\rm max} & \ge \sigma_0 + \rho_0 \sum_{h=1}^H L_h \,, \label{eq:Bnet-lower}\\
W_{\rm max} & \ge \frac{\sigma_0}{\min_h \{C_h\}} +  \sum_{h=1}^H L_h \,,
\label{eq:Wnet-lower}
\end{align} 
with 
\begin{align} 
\label{eq:L-h}
\!\!\!L_h = \min \left\{\frac{\sigma_h}{C_h - \rho_h} ,  
\frac{[\sigma_h + \rho_h [\Delta^h]_+ - C_h[\Delta^h]_-]_+}{C_h} \right\} . 
\end{align} 
\end{theorem}

\bigskip
\bigskip
The theorem implies that the backlog bound in 
Eq.~\eqref{eq:Bnet-simple} is sharp when 
$\Delta^h\ge 0$ for $h=1,\dots, H$, and, in particular, for FIFO.
For $\Delta^h<0$, observe that Eq.~\eqref{eq:L-h} implies that  
\[
0 \leq \theta^*_h - L_h \leq (1 - \frac{\rho_h}{C_h})[\Delta^h]_- \ . 
\]
Here, the deviation $\theta^*_h - L_h$  is due to  
the relaxation of the service curve in
Eq.~\eqref{eq-per-node-srv-crv-appx}.
The delay bound, although not sharp,
describes actual delays in adversarial scenarios rather
well, particularly on long paths.  We conjecture that an 
optimal choice of the $\theta_h$ 
would result in a delay bound that is as good as the backlog bound.

\begin{proof} 
For the arrivals at the first node, we set
$$
A_0(t)  =E_0(t)\,.
$$
We time the cross traffic to have a maximal burst
just before time $t=0$, if $\Delta^1\ge 0$,
and just before $t=\Delta^1$, if $\Delta^1<0$, 
resulting in 
$$
A^1_c(t)  = E_1(t+\nu -[\Delta^1]_-)\,,
$$
where $\nu>0$ is an infinitesimally small number. 

A careful analysis of the backlog from the cross flow at time $t=[\Delta^1]_+$, 
yields that the first bit of the through flow
experiences a latency of $L_1$. 
We can construct similar 
scenarios for cross traffic arrivals at subsequent nodes. For the $h$-th node, we time 
the arrival of a cross traffic burst relative to the
arrival of the first bit of through traffic,
by setting

$$
A_c^h(t)= E_h\bigl(t+\nu-\sum_{k=1}^{h-1}L_{k} -[\Delta^h]_-\bigr)\,. 
$$
Then,  the initial latency of the through flow 
is increased at the $h$-th node by~$L_h$. 
It follows that the first bit of the through flow
departs from the last node at time
$\sum_{h=1}^HL_h$, and so 

\vspace{-5pt}
$$
B_{\rm max}\ge E_0\bigl(\sum_{h=1}^H L_h\bigr)  \ ,  
$$
proving Eq.~\eqref{eq:Bnet-lower}. 
Now, consider the delay 
of the initial burst of the through flow. 
From the backlog analysis, we know that the first bit of the burst 
experiences a delay of $\sum_{h=1}^H L_h$. Since the rest  
of the burst cannot be transmitted faster than at the 
rate of the slowest link, we obtain Eq.~\eqref{eq:Wnet-lower}. 
\end{proof}

\section{Evaluation }
\label{sec:eval}

We present a numerical evaluation of the end-to-end bounds derived in 
this paper for deterministic and statistical arrival scenarios. 
We consider a network as in Fig.~\ref{fig:case-study}, where we assume for 
simplicity that all nodes have the same 
capacity of $C=100$~Mbps and use the same 
$\Delta$-scheduler.  We use a discrete-time setting with a step size 
of 1~ms.
The traffic arrivals of a flow follow an alternation of On and Off phases, 
where, in the  On phase, a flow transmits at rate $P=1.5$~Mbps, 
and the long term average arrival rate is $\rho=0.15$~Mbps. 
The total traffic load at each node is set to 600~flows with $N_0 = 10$ 
through flows and $N_c = 590$ cross flows, resulting in a link 
utilization of  $90$\%. 

For deterministic arrival scenarios, 
we characterize flows by a deterministic sample path envelope 
$E(t) = \rho t + \sigma$. We choose the value of $\sigma$  
such that the aggregate burstiness of 
all cross  flows is 300~Kb, and likewise for the through flows. 
(Selecting equal burst sizes for aggregate through and cross traffic 
provides a good separation of 
our upper and lower bounds for worst-case end-to-end delays.)  
Note that the deterministic scenario 
does not account for statistical multiplexing, i.e., the 
burst size of $N$ flows is $N$ times the burst of one flow. 

For statistical traffic, we assume that the arrivals of each flow are 
given by  
a discrete-time  Markov-Modulated On-Off process, with 
an On state ($=1$) and an Off state ($=0$).  
The probabilities to change  states are given by $p_{1\to0}=0.9$ and 
$p_{0\to1}=0.1$. 
The EBB characterization for the aggregate through and cross traffic can be obtained 
using the effective bandwidth $eb(\alpha)$ from 
(e.g., \cite{Book-Chang}, page 292),  
resulting in EBB parameters for the aggregate through traffic of 
$\rho_0= N_0 eb(\alpha_0)$, $M_0=1$, 
and $\alpha_0$ selected to achieve high statistical multiplexing gain. The parameters 
for 
the cross traffic are chosen accordingly. 

We assume that the through flows are independent of each other, 
and that the cross flows are independent; however, 
through and cross traffic is not independent. 
Thus, we observe statistical multiplexing for the group of through flows 
and the group of cross flows, but not across the groups. 

For the discussion of the examples, recall that $\Delta=0$ corresponds to FIFO, $\Delta=\infty$ to priority scheduling 
with low priority for through traffic,  and $\Delta=-\infty$ to priority scheduling 
with high priority for through traffic. 

\begin{figure}
\centerline{\includegraphics[width=0.7\columnwidth]{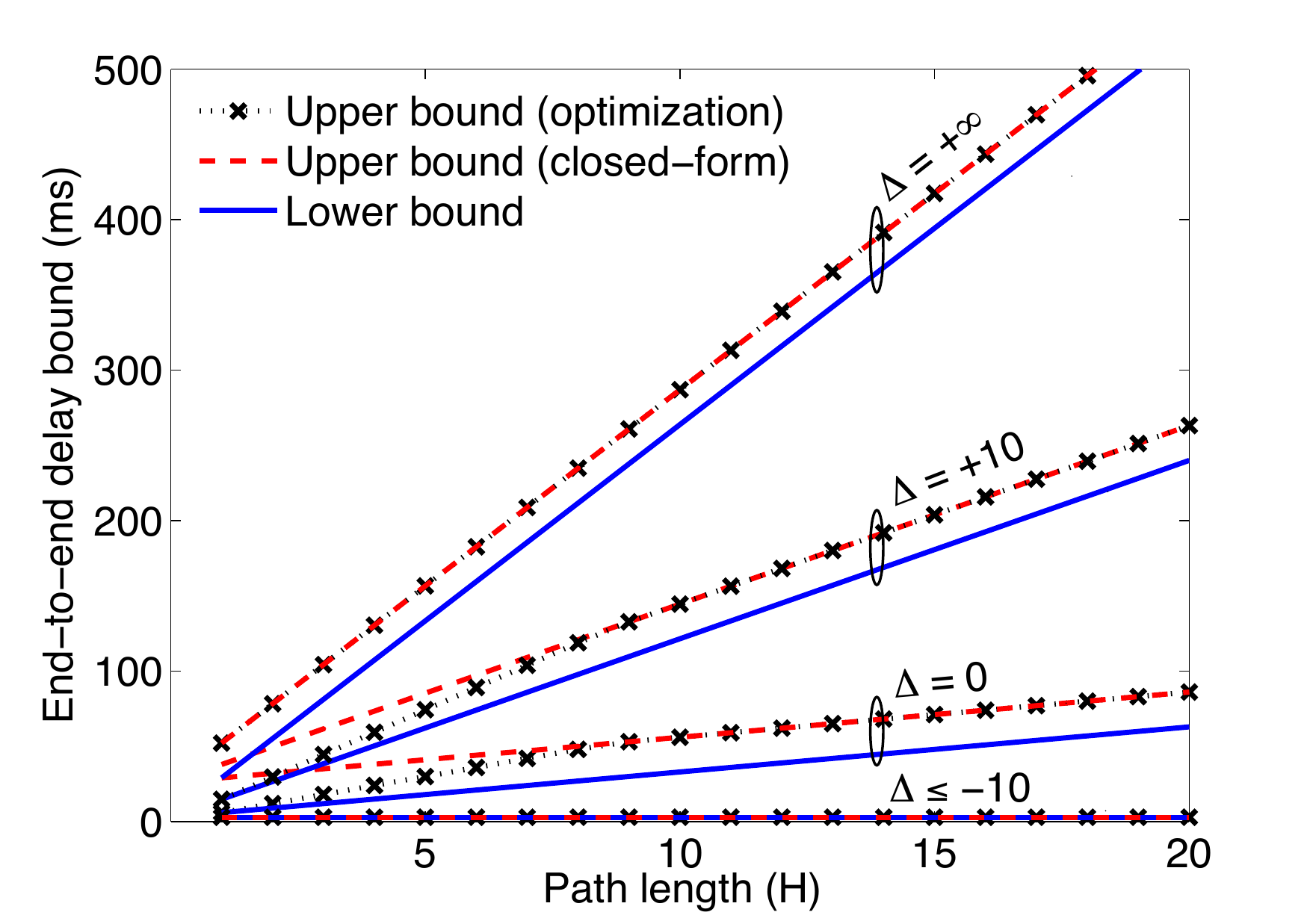}}
\caption{Deterministic end-to-end delay bounds of through traffic.}
\label{fig:Det_e2e_delay}
\end{figure}

\subsection{Tightness of Delay Bounds}

We evaluate the tightness of our delay bounds for the deterministic arrival 
scenario described above. We  compute the end-to-end 
delay bounds of the through traffic for different $\Delta$-schedulers as a function of the number of nodes~$H$. 
We present the closed-form delay bound $d^*_{\rm net}$ according to Eq.~\eqref{eq:Wnet-simple}, the bound computed with the optimization according to Eq.~\eqref{eq-opt-prob-1} 
(with $\tau_{\rm net}=\gamma=0$, since we are in a deterministic scenario), and 
the lower bound for the worst-case delay  from Eq.~\eqref{eq:Wnet-lower}.  

The results are shown in Fig.~\ref{fig:Det_e2e_delay}. 
The closed-form delay bounds track the 
bounds from the optimization very closely, and are virtually identical for long paths. 
Both upper bounds are highly accurate compared to the lower bound on the worst-case delay.  
Note that the differences between the delay bounds for different scheduling algorithms increase proportionally to the 
number of traversed nodes. Thus, in contrast to the scenario discussed in 
the introduction of the paper, we observe that the  impact of scheduling is just as strong 
on long paths as on a single node.

\begin{figure}
\centerline{\includegraphics[width=0.7\columnwidth]{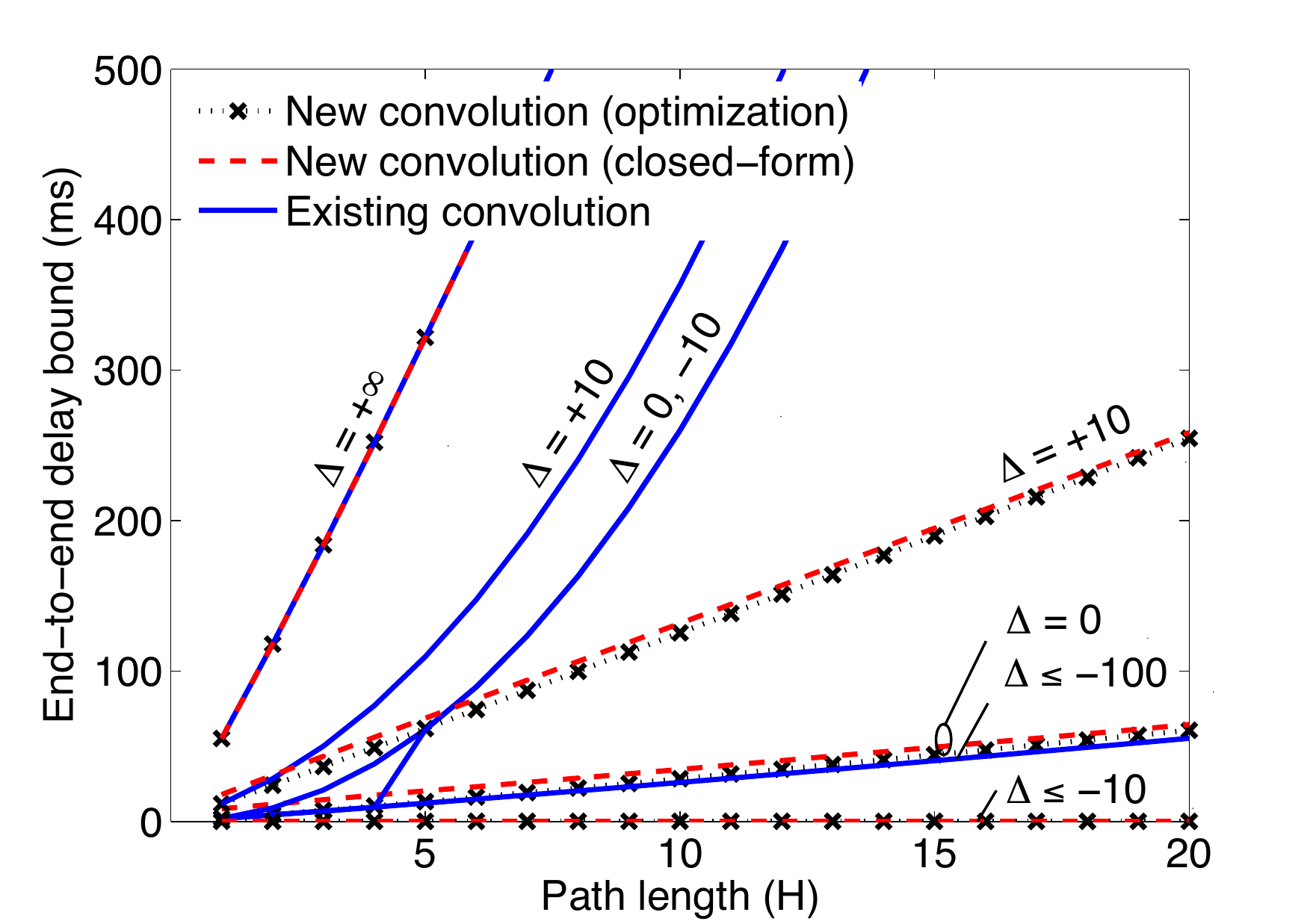}}
\caption{Statistical end-to-end delay bounds ($\eps = 10^{-9}$).}
\label{fig: Convolution}
\end{figure}

\subsection{Improvement due to Theorem~1}

For the remaining examples, we use the statistical traffic scenario. 
We first evaluate the benefit our convolution result, 
by comparing statistical delay bounds computed with the previously 
best convolution result from \cite{CiBuLi06}, to the delay bounds 
obtained with the new convolution result in Theorem~1. 
The delay bound computation for  $\Delta$-schedulers with the convolution result 
from \cite{CiBuLi06} uses an optimization, which is described in \cite{LiGhBu10}. 
For delay bounds computed with Theorem~1, we evaluate the closed-form  
expression from Theorem~\ref{thm:closed-form}, as well as the optimization in 
Eq.~\eqref{eq-opt-prob-1}. We use a fixed violation probability of 
$\eps = 10^{-9}$. 

In Fig.~\ref{fig: Convolution}, we show the end-to-end delay bounds  of the through traffic for different $\Delta$-schedulers for increasing path length. 
In all cases, the closed-form delay bounds are very close to 
the bounds obtained by the optimization. For $\Delta=\infty$, i.e., a priority scheduler where 
through traffic has lowest priority, the delay bounds obtained with the new convolution 
theorem  give the same results as  \cite{CiBuLi06}. 
Otherwise, the new convolution theorem results in dramatically 
improved delay bounds, especially for long paths. 
Our results clearly motivate 
the need for our new statistical network service curve expression 
for schedulers with $\Delta < \infty$. 
In fact, the delay bounds obtained with the `old' convolution method for service curves 
show a similar growth rate for large $H$, 
thus, systematically underestimating the difference between schedulers. 

\begin{figure}
\centerline{\includegraphics[width=0.7\columnwidth]{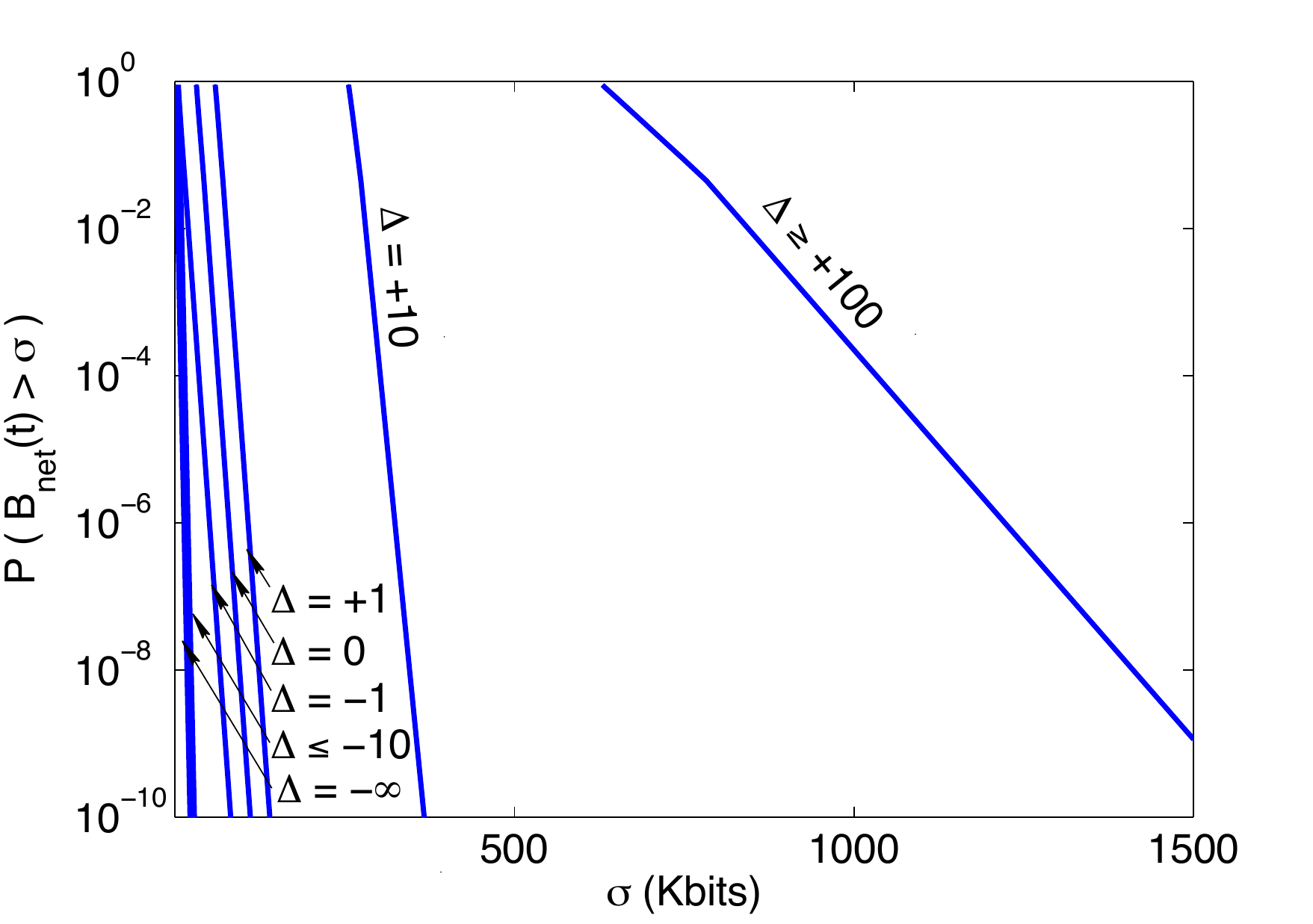}}
\caption{Tail bound of backlog distribution ($H=10$).}
\label{fig:Backlog_violationP}

\centerline{\includegraphics[width=0.7\columnwidth]{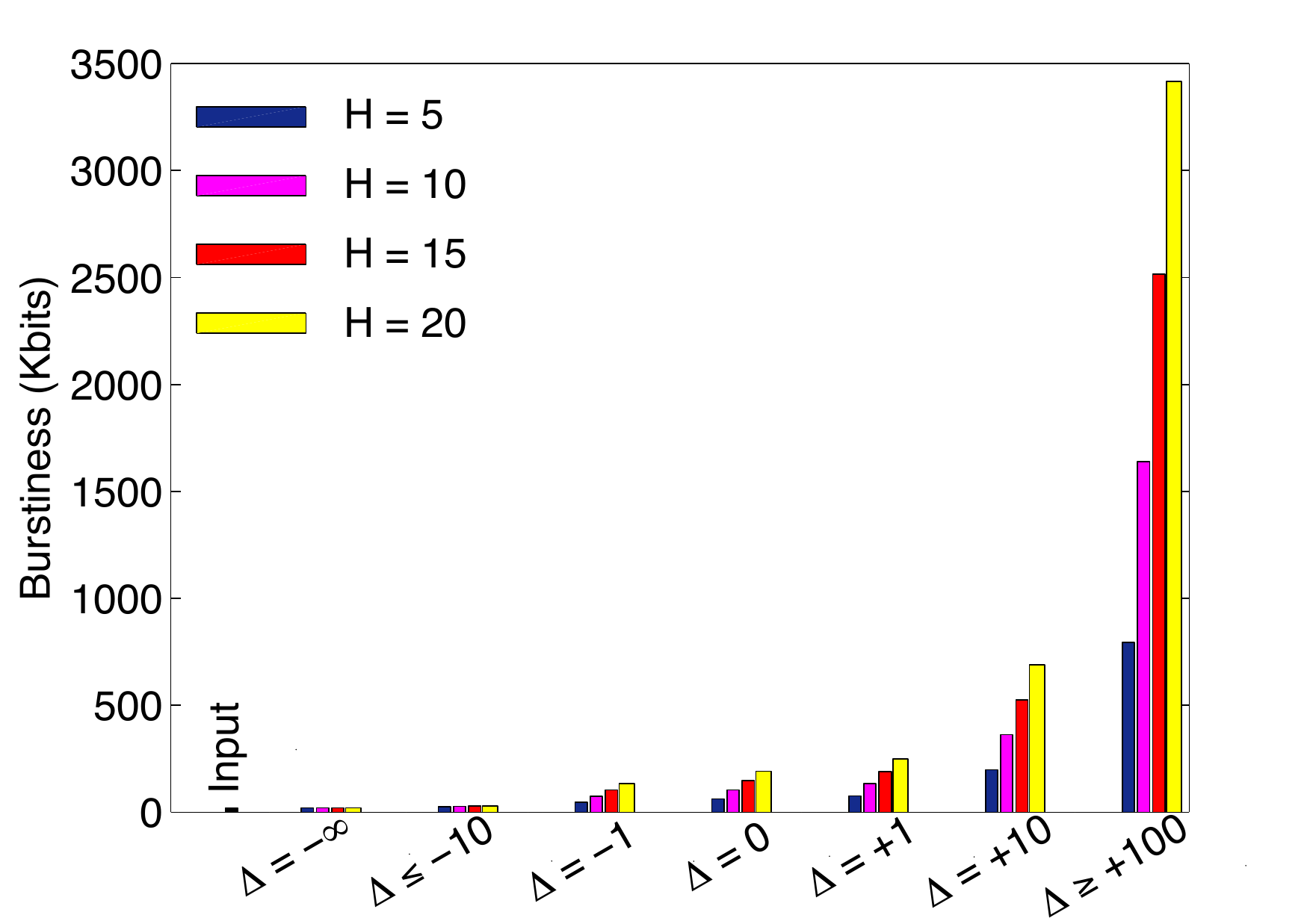}}
\caption{Burst size of output traffic ($\eps = 10^{-9}$).}
\label{fig:Decompose}

\vspace{-5mm}
\end{figure}

\subsection{Tail Bound of Backlog Distribution}

Fig.~\ref{fig:Backlog_violationP} shows the tail distribution for the backlog 
of the through traffic for a path of $H=10$~nodes. The graph shows 
that the backlog distribution divides  the $\Delta$-schedulers into 
two groups:  
one for  large values of $\Delta$, and one for 
small and negative values of $\Delta$. In the first group, through traffic 
is treated 
effectively as low-priority traffic. 
The second group has a backlog distribution similar to  FIFO ($\Delta=0$),  
indicating that this desirable property of FIFO scheduling 
extends to some non-FIFO algorithms.

\subsection{Burstiness Increase of Output Traffic}

Studying the burstiness of the output traffic  further corroborates 
the findings of the previous example. 
Here, we compute bounds for the output  of the through traffic 
for different choices of $\Delta$ and $H$, by applying the closed-form 
bound for $D_{\rm net}$ from Theorem~2. 

In Fig.~\ref{fig:Decompose} 
we show the burst size of the output bound 
for a violation probability $\eps=10^{-9}$ in a bar chart. 
(The rate values are omitted since the long-term traffic rate does 
not change at the output of a link.)  
For comparison, the leftmost bar shows the burst of the through traffic 
arrivals at the first node.  
For large values of $\Delta$, we see that  
the burstiness increases sharply. 
For values of $\Delta$ close to zero or negative, we observe only a modest 
increase of the burstiness at the output, even when the number of nodes is large. 
As a result, the 
arrival process at downstream nodes is not markedly different from that 
at the first node. 
Since modest increases of output burstiness are  
an important condition for a network decomposition analysis, e.g., see  
\cite{Eun05,Mazum05}, our work suggests that conditions 
for decomposition may not be limited to FIFO networks.

\section{Conclusions}
\label{sec:conclusions}

The analytical methods and closed-form expression developed in 
this paper vastly enhance available methods for end-to-end performance analysis 
in networks with deterministic and statistical traffic. 
With our analysis, it is now possible to make conclusive 
statements about the role of link scheduling in 
large networks. Contrary to conventional wisdom, 
we showed that the impact of link scheduling generally does not 
diminish on long network paths. We also shed light 
on a group of scheduling algorithms that provide service differentiation 
without the penalties of strict priority scheduling in terms 
of buffer requirements and deterioration of the burstiness of traffic.

\end{document}